\documentclass[10pt,letterpaper]{article}
\usepackage{opex3}

\usepackage{cite}

\begin{document}


\title{A versatile integrating sphere based photoacoustic sensor for trace gas monitoring}

\author{Mikael Lassen,$^{1,*}$ David Balslev-Clausen,$^1$ Anders Brusch,$^1$ and Jan C. Petersen$^1$}

\address{$^1$Danish Fundamental Metrology, Matematiktorvet 307, DK-2800 Kgs. Lyngby, Denmark}

\email{$^*$ml@dfm.dk} 



\begin{abstract}
A compact versatile photoacoustic (PA) sensor for trace gas detection is reported. The sensor is based on an integrating sphere as the PA absorption cell with an organ pipe tube attached to increase the sensitivity of the PA sensor. The versatility and enhancement of the sensitivity of the PA signal is investigated by monitoring specific ro-vibrational lines of CO$_2$ in the 2 $\mu$m wavelength region and of NO$_2$ in the 405 nm region. The measured enhancement factor of the PA signal exceeds 1200, which is due to the acoustic resonance of the tube and the absorption enhancement of the integrating sphere relatively to a non-resonant single pass cell. It is observed that the background absorption signals are highly attenuated due to the thermal conduction and diffusion effects in the polytetrafluoroethylene cell walls. This demonstrates that careful choice of cell wall materials can be highly beneficial to the sensitivity of the PA sensor. These properties makes the sensor suitable for various practical sensor applications in the ultraviolet (UV) to the near infrared (NIR) wavelength region, including climate, environmental and industrial monitoring.
\end{abstract}




\section{Introduction}

Versatile, highly sensitive, low cost and easy to operate trace gas detection systems are important for a number of practical applications, including climate, environmental and industrial monitoring. Trace gas measurements in the MIR and NIR wavelength regions are particular important due to the presence of strong ro-vibrational bands of most molecules resulting in high sensitivity \cite{Sigrist1994,Sigrist2008,Patel2008}. Photoacoustic spectroscopy (PAS) is a very promising method due to its ease of use, relatively low cost and the capability of allowing trace gas measurements at the sub-ppb level \cite{Harren2000,Nägele2000,Miklos2001,Xu2006,Michaelian2003,Koskinen2008,Besson2006}. These outstanding features of PAS can only be fully exploited using a suitable designed acoustic resonator in combination with a high power light source due to the fact that sensitivity is proportional to the optical power and the acoustic enhancement factor \cite{Rosencwaig1980Book}. The modulation frequency of the light source and thus the acoustic wave needs to be matched to the resonance frequency of the acoustic cell, resulting in a PA signal amplified by the acoustic quality factor $Q$. High power and tunable light sources tend to be big and bulky and limit the compactness of the spectrometer \cite{Webber2003}. However, the PA signal may also be enhanced by optical multi-pass techniques resulting in an increase of the sensitivity of the PA spectrometer due to the increased light absorption path length from multiple reflection. Various multi-pass and single-pass configurations have so far been exploited for PAS configurations, such as ring cells, cavity based cells and transverse square cells  \cite{Nägele2000,Rey2005,Miklos2006,Saarela2010,Manninen2012}.  However, most existing technologies with small and compact size have relatively low sensitivity and limited spectral resolution. Therefore, novel solutions are desirable. The use of an integrating sphere as absorption cell simplifies the sensor since the optical alignment is very simple and multiple passes automatically are achieved. The sensor can be made compact and highly versatile. The integrating sphere based PA sensor is in principle suited for many different detection techniques and has so far mostly been used for direct absorption including various modulation techniques \cite{Elterman1970,Hodgkinson2009,Tranchart1996,Hawe2005}.

The scope of this paper is to demonstrate the signal enhancement and the versatilely of the combination of an integrating sphere as multi-pass absorption cell and an attached organ pipe tube as a coupled acoustic resonator. A PA sensor based on an integrating sphere manufactured from polytetrafluoroethylene (PTFE) is presented. The PTFE material is a highly reflective in the wavelength range 250 - 2500 nm (UV - NIR). In this region the average reflectivity of PTFE is higher than $>95\%$. However due to the uniform distribution of the light field inside the integrating sphere acoustic resonances can not be exploited directly \cite{Rosencwaig1980Book}. It is demonstrated that the sensitivity of the PA sensor can be increased by attaching a 90 mm long organ pipe tube to the integrating sphere thereby making use of the acoustic resonance of the tube. The versatility and enhancement of the PAS sensitivity of the integrating sphere has been investigated by monitoring specific ro-vibrational lines of CO$_2$ in the 2 $\mu$m wavelength region and of NO$_2$ in the 405 nm wavelength region \cite{Lewicki2007,Hawe2006,Bernhardt2010,Yi2011}. The absolute sensitivity of the system is estimated to a minimum single-shot detectable CO$_2$ concentration in the 2 $\mu$m region of approximately 30 ppm and a minimum detectable NO$_2$ concentration of 1.9 ppm at a SNR = 1. These two gases are important in climate and environmental monitoring. NO$_2$ is a toxic atmospheric pollutant and is mainly emitted into the atmosphere due to combustion processes. The average mixing ratio of NO$_2$ in the atmosphere is typically between 5-30 parts per billion, however close to a combustion engine it can be orders of magnitude higher. Enhancements exceeding 1200 of the PA signal were observed. It is observed that the background absorption signals are highly attenuated due to the thermal conduction and diffusion effects in the PTFE cell walls. This suggests that the background signal issue of typical PA measurements can be circumvented by appropriate choice of cell wall materials in additional to careful optical alignment to ensure background free PA measurements. This makes the sensor suitable for many practical applications in the UV - MIR wavelength range, including climate, environmental and industrial monitoring of trace gasses.

\section{Theoretical considerations}

The PA effect was first reported in 1881 by Bell \cite{PAS1881}, however, it was not until after the invention of the laser that the effect found use in sensitive spectroscopy in the 1970s \cite{Tam1986}.
The PA technique is based on the detection of sound waves that are generated due to absorption of modulated optical radiation. A microphone is used to monitor the acoustic waves that appear after the laser radiation is absorbed and converted to local heating via molecular collisions and de-excitation in the PA cell \cite{Demtroder2003}. The magnitude of the measured PA signal is given by \cite{Rosencwaig1980Book}:
\begin{equation}
 S_{PA} = S_m P F \alpha,
 \label{eq.PAsignal}
\end{equation}
where $P$ is the power of the incident radiation, $\alpha$ is the absorption coefficient, which depends on the the total number of molecules per cm$^3$ and the absorption cross section, $S_m$ is the sensitivity of the microphone and $F$ is the cell-specific constant, which depends on the geometry of acoustic cell and the quality factor $Q$ of the acoustic resonance \cite{Sigrist1994,Harren1997}. Since according to Eq.(\ref{eq.PAsignal}) the PA signal is proportional to the density of molecules the technique is able to measure absorption directly, rather than deriving it from the transmission spectrum. However, the PAS technique is not a metrological absolute technique and requires calibration using a certified gas reference sample. Ideally, a highly sensitive PA system should amplify the sound wave and reject acoustic and electrical noise as well as in-phase background absorption signals from other materials in the cell (walls and windows). Eq.~(\ref{eq.PAsignal}) shows that the sensitivity of the PA signal increases with laser power. Higher sensitivity can thus be achieved either by increased laser power or by overlapping laser beams multiple times through the gas sample. Various multi-pass and single-pass configurations have sofar been exploited for PAS, including standard single pass cylindrical cells, ring cells, cavity based and transverse square based  \cite{Miklos2006,Saarela2010}.

\subsection{Integrating sphere as absorption cell}

Using an integrating sphere as absorption cell simplifies the system. Integrating spheres are typically used for a variety of optical, photometric or radiometric measurements. For integrating spheres the measure for optical intensity enhancement is the so called sphere multiplier $M$ that accounts for the increase in radiance due to multiple reflections in the sphere. The multiplier is simply a function of the average reflectance, $\bar{\rho}$, of the sphere and the initial reflectance for incident radiant, $\rho_0$, and is given by:
\begin{equation}
M = \frac{\rho_0}{1-\bar{\rho}}\label{eq.Q multiplier}.
\end{equation}

The PTFE sphere used in this work has an average reflectance $\bar{\rho} = 0.95$ at 2 $\mu$m while at 405 nm it is $\bar{\rho} = 0.98$, where we have taken the apertures of the integrating sphere into account. This results in multiplier values of 20 and 50, respectively. Thus enhancing the optical intensity and hereby the sensitivity of the sensor by these factors. The total enhancement of the system relative to a single-pass non-resonant PAS is therefore: $ S_{enhancement}= Q \times M$.

\begin{figure}[t]
\centering\includegraphics[width=10cm]{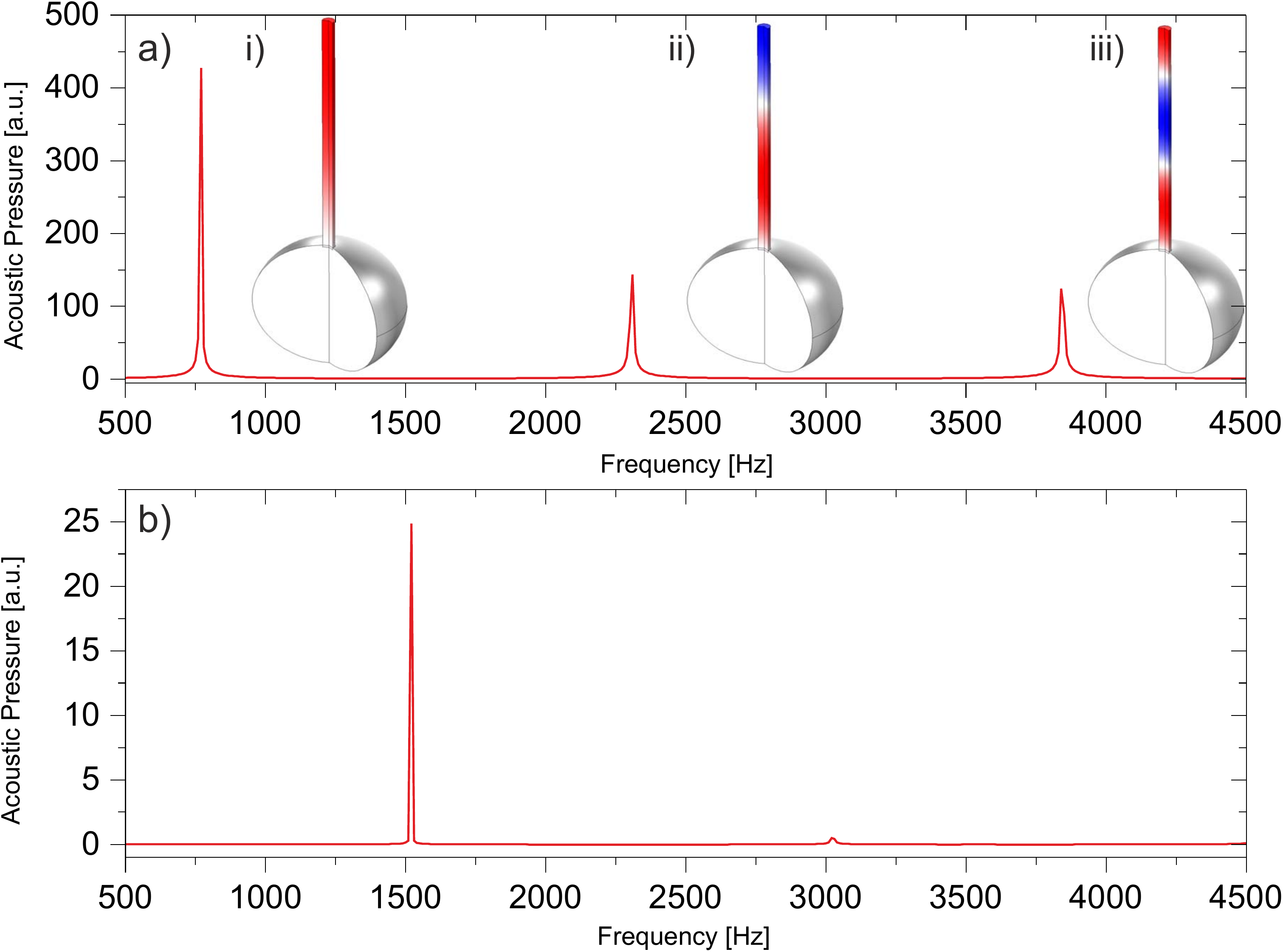}
\caption{Simulations of the acoustic coupled system. (a) Acoustics pressure response as a function of frequency for the organ pipe tube. (b) Acoustics pressure response as a function of frequency for the sphere. The figure shows the 3D simulation of first three eigenfrequencies of the coupled sphere and cylindrical acoustic resonator measured at end of the tube.  i) 743 Hz, ii) 2229 Hz and iii) 3716 Hz. The blue and red colors indicate the maximum and minimum acoustic pressure, with opposite phase. The white color is zero acoustic pressure. The corresponding experimental data is shown in Fig.~\ref{NO2_sensor}.}
\label{comsolsimulations}
\end{figure}

In a typical PA systems the PA signal is enhanced by the acoustic resonances of the absorption cell. Unfortunately the acoustic resonance cannot be exploited directly in the present setup due to the uniform distribution of the light field inside the integrating sphere. This can be seen from the overlap integral between the acoustic pressure field, $p_j$, and the optical intensity distribution $I(r,\omega)$. For a spatially constant optical field distribution we see \cite{Rosencwaig1980Book} that for all $j$ except for $j=0$ which is the only non-zero solution with an acoustic frequency of $\omega_0=0$. In order to overcome the constrains of the non-resonant integrating sphere and exploit the acoustic resonances as an enhancement factor a 90 mm long organ pipe tube is attached to the integrating sphere, thus allowing for an enhancement of the PAS signal due to the acoustic resonance of the tube. Simulations of the coupled acoustic system (sphere and organ pipe) were performed with a 3-dimensional model using a finite element model (FEM) multi-physics simulation program COMSOL. The pressure acoustic module is used to solve the Helmholtz equation. The boundary conditions were hard walls. The acoustic field is excited by applying a uniform pressure on the sphere walls. The simulation conditions were 1 atm pressure and at 25 $^\circ$C and a speed of sound of 343 m/s. The simulations are shown in Fig.~\ref{comsolsimulations} and shows the theoretical simulated frequency response of the coupled acoustic system. The tube material is aluminum and the length and radius are 90 mm and 2 mm, respectively. The integrating sphere absorption cell has a radius of 25.4 mm and the material is PTFE. These are the same conditions as for the experimental realization. When the integrating sphere is excited by an uniform pressure wave, no resonance is observed in the integrating sphere, however, acoustic resonances in the organ pipe are excited. In Fig.~\ref{comsolsimulations}(a) the first three eigenmodes of the coupled acoustic system are shown and depicted as if a microphone were attached to the end of the organ pipe tube. The eigenfrequencies of the modes are at i) 743 Hz, ii) 2229 Hz and iii) 3716 Hz, respectively. Note that the first eigenfrequency is relatively low and for real practical applications this could pose a potential problem since acoustic noise from the surroundings would interfere with the PA signal. This can be circumvented by choosing a shorter organ pipe tube, thus moving the resonance toward higher eigenfrequencies. Note that in a typical experiment the light field is not necessarily completely uniform and acoustic resonances may be excited in the integrating sphere. From the simulations this can be realised by exciting the integrating sphere with a nonuniform pressure wave. The top figure in Fig.~\ref{comsolsimulations}(b) shows the theoretical simulated frequency response of sphere as if a microphone were attached to the equator of the sphere, where a response at 1500 Hz and 3000 Hz is found. Note that in pure CO$_2$ and the speed of sound is approximately 267 m/s at 1 atm and 20 $^\circ$C, thus it is expected that the fundamental resonance is shifted to approximately 650 Hz for pure CO$_2$.

\section{Experimental setup}

\begin{figure}[t]
\centering\includegraphics[width=10cm]{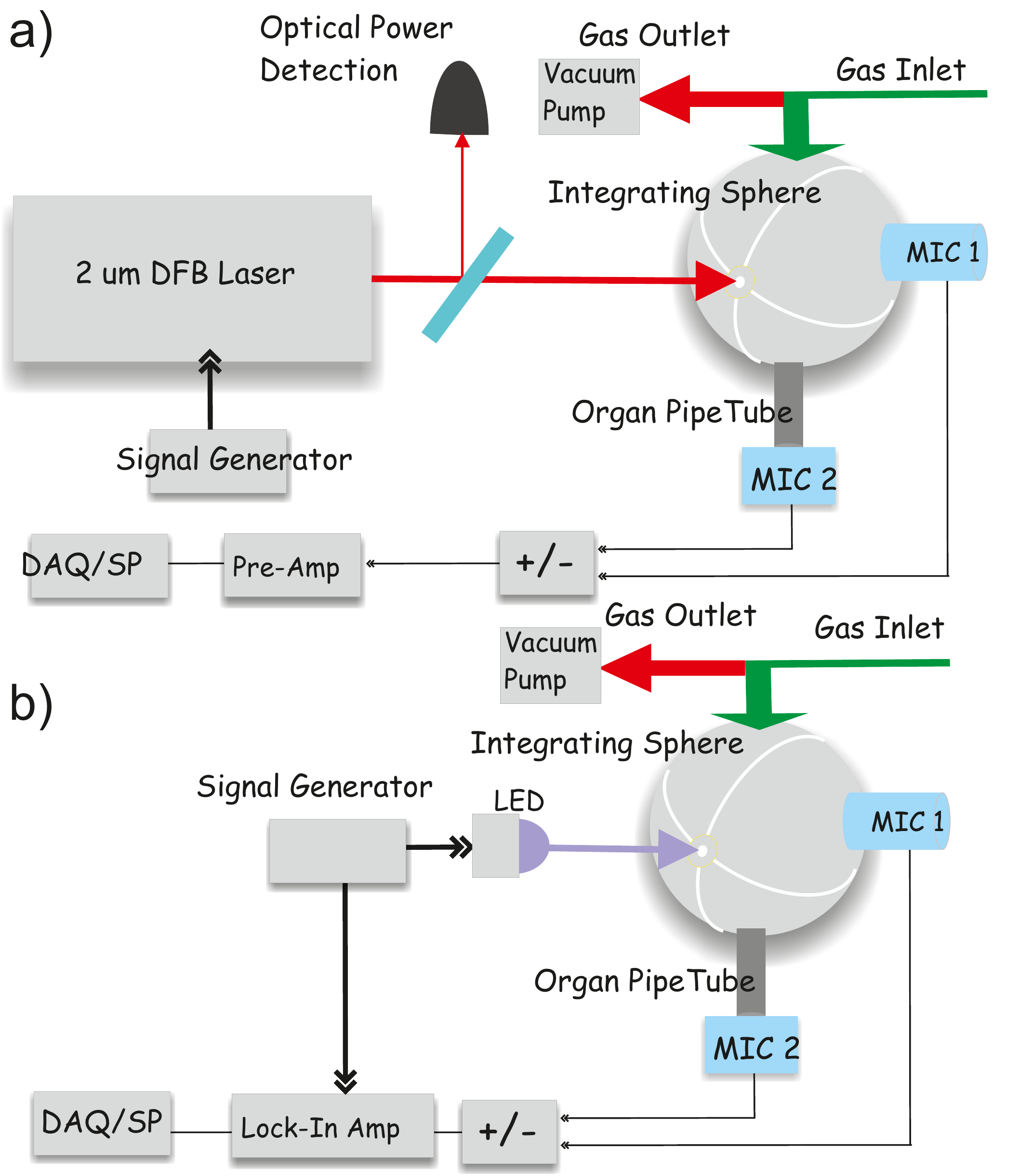}
\caption{(a) The experimental setup for CO$_2$ monitoring consist of a distributed-feedback (DFB) diode laser emitting radiation at 2.004 $\mu$m, an integrating sphere with a diameter of 50.8 mm and two microphones attached to the integrating sphere. One directly on the integrating sphere and one attached via the 90 mm organ tube pipe. DAQ is a data acquisition card. (b) The setup for measuring NO$_2$ include a 405 nm LED source and a lock-in amplifier connected to a DAQ card.}
\label{experimentsetup}
\end{figure}

A typical experimental setup for PAS involves a light source, that is either mechanically chopped or current modulated, and an absorption cell with microphones or a pressure sensitive detector. Two experimental setups were used, one using a DFB laser light source for monitoring specific ro-vibrational lines of CO$_2$ in the 2 $\mu$m region and one using a LED light source for investigating NO$_2$ in UV/Blue region \cite{Demtroder2003}. The schematics of these are shown in Fig.~\ref{experimentsetup}. The laser based setup is shown in Fig.~\ref{experimentsetup}(a). The DFB laser emits light at 2.004 $\mu$m ($\pm 0.002 \mu$m) and is used to probe the R(12) line of CO$_2$. The laser wavelength was fine tuned by changing the temperature allowing for a change of 0.26 nm/C. The 50.8 mm diameter integrating sphere was manufactured from polytetrafluoroethylene (PTFE), which is a high reflective bulk material in the wavelength range 250 - 2500 nm (UV - NIR). The reflectivity in this region is higher than $>95\%$, resulting in a mean light path length of approximately 1.2 m. The light beam enters the integrating sphere via an uncoated 3 mm thick calcium fluoride window. The optical transmission is 92$\%$ in the wavelength region 0.2-8 $\mu$m and the absorption coefficient is 10$^{-4}$/cm. The laser beam then hits the cell wall opposite of the window and is scattered so that the light field is evenly distributed over all angles. Due to the placement of the window outside the sphere the background signal is decoupled from the PA signal and was not detectable. Two microphones were attached to the integrating sphere. One directly on the sphere and one at the end of the organ tube pipe. The measurements were performed at the experimental conditions of 1 atm and at 20 $^\circ$C. The optical modulation was approximately 1.4 mW peak-to-peak. The amplitude modulation of the optical field is made by switching the laser current on and off, thus generating the PA signal at this particular modulation frequency. However, pure amplitude modulation is not easily achieved, as residual wavelength modulation also occurs. Alternatively, the PA sensor could also be operated in the wavelength modulation mode in which case the PA signal would be excited at the overtone frequency \cite{Schilt2006}. The data are collected with a DAQ card having a bandwidth of 9 kHz at 50 ks/S and 12 bit resolution.

The schematics of the experimental setup for the NO$_2$ measurements is shown in Fig.~\ref{experimentsetup}(b), where the DFB laser has been substituted with a 405 nm LED having a 13 nm bandwidth (FWHM). The NO$_2$ molecule has a strong and broad absorption spectrum covering the 250-650 nm spectral, however, below 415 nm photochemical dissociation of NO$_2$ occurs \cite{Pitts1964,Barreiro2010}. Above 415 nm approximately 90$\%$ of the absorbed light is  converted to heat/pressure through the PA effect. The data is processed using a lock-in amplifier. The LED modulation is controlled by a signal generator, which also acts as the local oscillator for the lock-in amplifier. The peak-to-peak modulation is 130 mW, and approximately 80 mW is coupled into the sphere. The data from the lock-in amplifier is collected with a DAQ card with a bandwidth of 9 kHz at 256 ks/S and 16 bit resolution.

\subsection{Results}

\begin{figure}[t]
\centering\includegraphics[width=10cm]{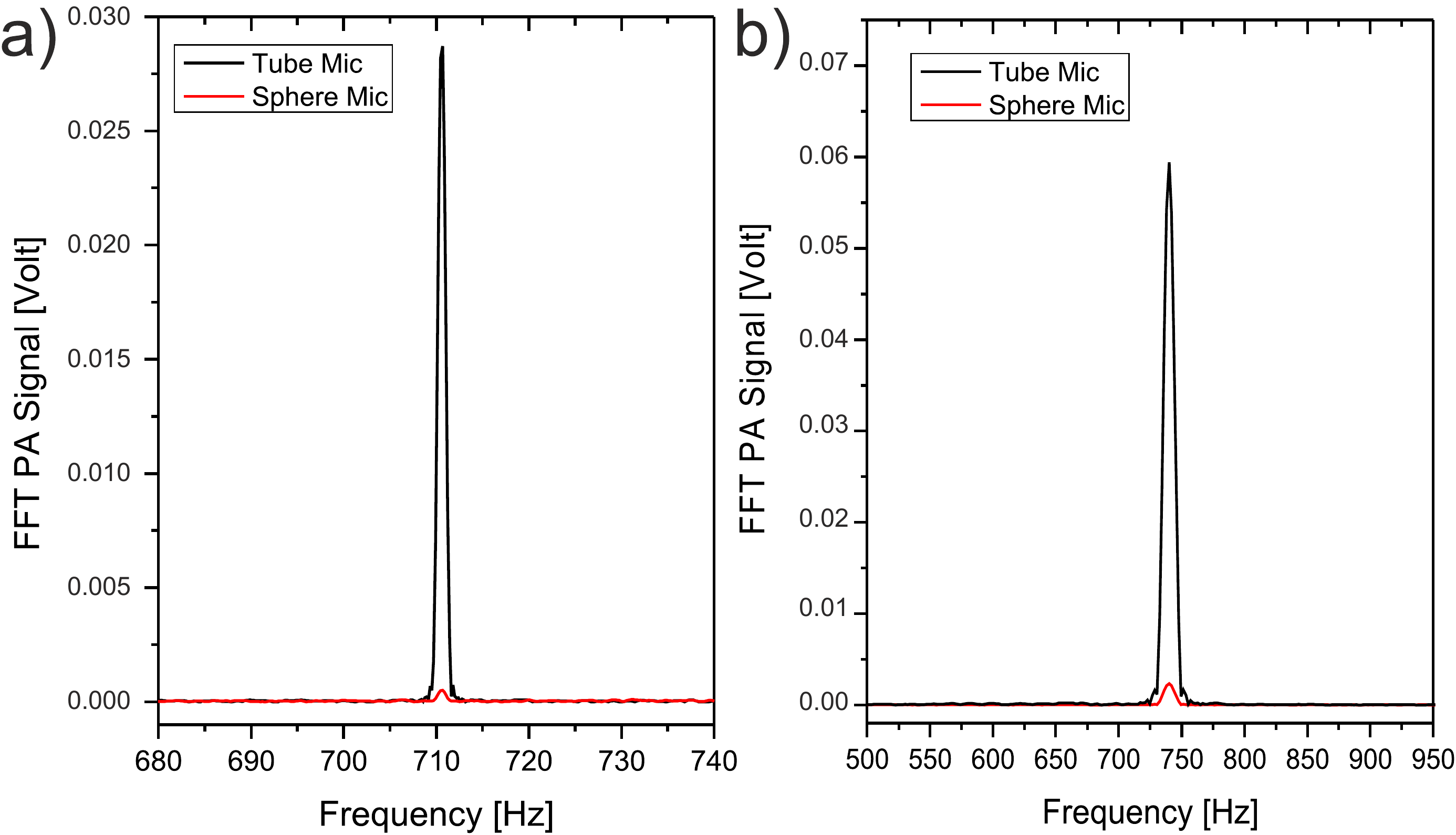}
\caption{Enhancement factors of the integrating sphere based PA sensor for (a) CO$_2$ and (b) NO$_2$. The black curves are the data recorded by Mic2 at the end of the organ pipe while the red curves are the data recorded by Mic1, situated inside the sphere. The enhancement factors for the two cases are indicated.}\label{DataTwoMics}
\end{figure}

Fig.~\ref{DataTwoMics} shows the enhancements of the PA signal due to the acoustic resonance of the organ pipe tube. The PA signals from the two microphones are shown; one at the end of the organ pipe (Mic2, black curve) and one directly attached to the sphere (Mic1, red curve). The latter provides the non-resonant signal. The signal shown in Fig.~\ref{DataTwoMics}(a) is due to a pure CO$_2$ filled sphere. The PA signal (sensitivity) is enhanced by factor of approximately 58 compared with the non-resonant signal. It can be concluded that the Q-factor of the organ pipe is approximately 58. The signal shown in Fig.~\ref{DataTwoMics}(b) is due to a 300 ppm NO$_2$ in N$_2$ filled sphere. The PA signal is enhanced by a factor of approximately 30. However this enhancement factor is probably slightly higher if the background absorption is taken into account and also the interference from ambient noise. The enhancement due to the long path length provided by the sphere compared with a single pass absorption cell can be estimated from Eq.~(\ref{eq.Q multiplier}) to be approximately 20 and 50 for the CO$_2$ and NO$_2$ setup, respectively. The total enhancement factor for the CO$_2$ and NO$_2$ measurements is approximately 1200 and 1500, respectively, compared with a simple single pass non-resonant cell with the same incident optical power.

The SNR (PA signal over the background signal with no light or SBR) of the high concentration CO$_2$ measurements is approximately 4500. Note that in using pure CO$_2$ the PA signal is saturated and therefore the SNR would be higher for pure CO$_2$ if there were no saturation effects. It is observed that by diluting the pure CO$_2$ with atmospherical air so that the CO$_2$ concentration is lowered to 20$\%$ we see that the PA signal size does not decrease.  The saturation of the PA signal occurs when the pump becomes depleted due to a highly dense gas, thus all light is therefore absorbed in the first 20-30 cm of the integrating sphere. The mean light path length of the sphere is approximately 1.2 m this gives therefore the same signal for a 20$\%$ CO$_2$ mixture with atmospheric air as for the pure CO$_2$. The pressure is kept at atmospheric pressure at all times. The absolute concentration and thus the sensitivity of the sensor is difficult to estimate from this high concentration (pure) CO$_2$. By observing the shift in the resonance frequency due to the change in the speed of sound, which is 267 m/s and 343 m/s for pure CO$_2$ and atmospherical air, respectively, it is estimated that the concentration is approximately 15$\%$ CO$_2$. With this SNR the minimum detectable CO$_2$ concentration in the 2 $\mu$m region is approximately 30 ppm (SNR =1) for a single-shot. We would like to point out that the scope of this paper is to demonstrate the signal enhancement due to the integrating sphere as simple kind of multi-pass absorption cell and the further enhancement of the signal due to the attached organ pipe tube and not to demonstrate an absolute sensitivity of the system. However, the sensitivity can easily be enhanced by using higher optical power, better microphones and longer integrating time, which would make the system comparable to state-of-art PA sensors.

\begin{figure}[t]
\centering\includegraphics[width=10cm]{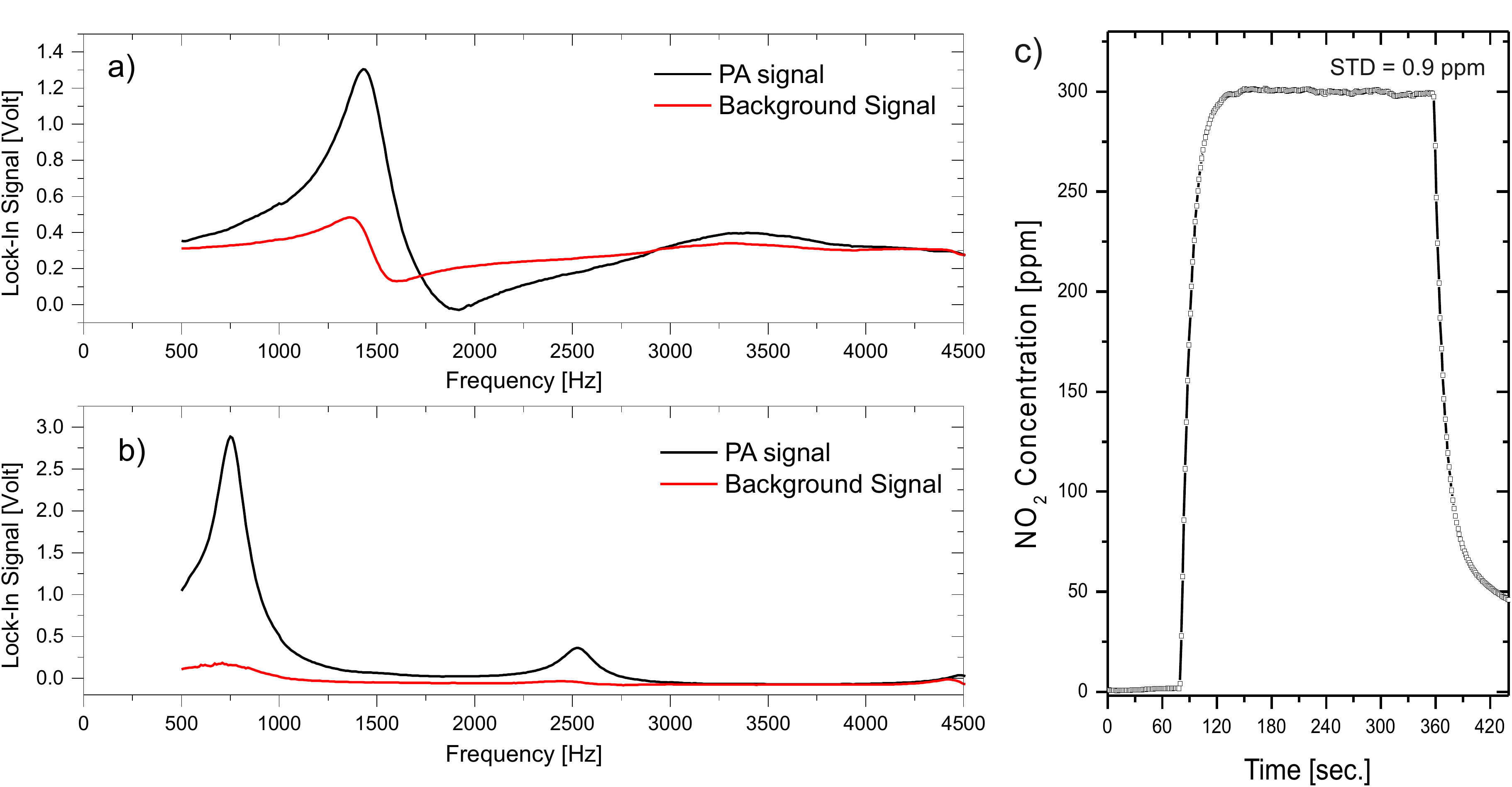}
\caption{Data for measurement on a 300 ppm NO$_2$ mixture, SNR = 22 dB. (a) Sphere microphone PA signal (black curve) and background signal (red curve). (b) Tube microphone PA signal (black curve) and background signal (red curve) The eigenresonances are found at approximately 740 Hz and 2500 Hz for the tube microphone. (c) Monitoring of a 300 ppm NO$_2$ concentration over 3.5 minutes resulting in a standard deviation of 0.9 ppm. }\label{NO2_sensor}
\end{figure}

The results of the NO$_2$ measurements are shown in Fig.~\ref{NO2_sensor}. The measurements are processed with a lock-in amplifier with a integration time of 1 second. Fig.~\ref{NO2_sensor} (a) and (b) show the frequency response, for the PA signals and background signals of the microphones attached to the integrating sphere and the organ pipe tube, respectively. It is observed that the SNR is approximately 160 for the organ pipe tube signal at a modulation frequency of 740 Hz. The minimum detectable NO$_2$ concentration is therefore approximately 1.9 ppm (SNR =1). The largest contribution to the background seems to originate from external ambient noise sources, such as vacuum pumps and therefore the SNR is probably considerably higher (500 or more). A SNR of 500 corresponds to a minimum detectable absorption coefficient of 3.8 $\cdot 10^{-7}$cm$^{-1}$ Hz$^{1/2}$ W$^{-1}$ which is comparable to state-of-the-art PA NO$_2$ measurements \cite{Saarela2011}. The SNR for the PA signal measured with the sphere microphone is approximately 4 measured at 1500 Hz, thus the minimum detectable NO$_2$ concentration is approximately 75 ppm (SNR =1). This relatively high background signal is due to the fact the microphone is placed inside the sphere and is being heated by the incident light. On the other hand the tube microphone is protected from the stray light by placing the microphone at the far end of the tube. The resonances of the tube can be observed at 740 Hz and 2500 Hz as expected from the simulations discussed above. The sphere has a resonances at approximately at 1500 Hz which was in good agreement with the above discussed simulation.  Note that the size of the total acoustic pressure strength and hence the Q-factor of the peaks in the simulation does not agree very well with the experiment. This is attributed to the fact that our simulation does not take damping and general loss factors into account. Fig.~\ref{NO2_sensor}(c) shows results due to a continuously flow of 300 ppm NO$_2$ through the PA cell resulting in a standard deviation (STD) of 0.9 ppm for measurements during 3.5 minutes.  The flow was maintained in order to keep the concentration at the same level while photo-dissociation of the NO$_2$ molecule was active. It is expected that the STD will be lower by a factor of 10 if photo-dissociation were not present. By stopping the flow of NO$_2$ into the sphere the photochemical dissociation of the molecules becomes apparent and within 2 minutes the concentration is lower by a factor 10.

\subsection{Background Signal}

These relatively high SNR/SBRs of the PA signals are slightly surprising since the mean reflectivity is only 95$\%$(98$\%$) for the 2 $\mu$m (405 nm) measurements and in the case of an empty cell (no absorbance by gas) all incident light is absorbed in the cell walls (PTFE material). In standard PAS experiments such a high level of background absorption would be detrimental to the performance of the system.  This surprising features is due to a combination of mechanisms, namely the large light penetration depth, the low thermal diffusion length and heat conduction to the outer aluminium casing of the integrating sphere. The light penetration depth is around 1.39 mm at a wavelength of 633 nm \cite{Sidorov2012}. The diffusion length scale for pulsed heating is given by \cite{Bird1976}: $\mu_t = 2 \sqrt{\alpha t}$, where $\alpha$ is the thermal diffusivity, which is 0.124 mm$^2/s$ for PTFE at 25 $^\circ$C and $t$ is the pulse duration time. In our experiment the modulation frequency is around 700 Hz which leads to a diffusion length of approximately 27 $\mu$m. Thus the light that penetrates the PTFE material deeper than 27 $\mu$m only contributes to the background absorption signal with a constant DC heating component and does not contribute to an in-phase PA signal at 700 Hz. The amount of absorbed light that contributes to the background signal is given by the ration between the diffusion length and the penetration depth and is on the order of 2$\%$. In our CO$_2$ experiment the average optical power is 2 mW, frequency of 700 Hz and with a 50-50 duty circle. This would lead to a background signal of approximately 20 ppb. In comparison if we consider aluminum as the cell material. Aluminum has a thermal diffusivity of 84 mm$^2/s$, which leads to a diffusion length of 700 $\mu$m, however since the optical penetrations depths is only a few nm. Therefore diffusion effects will be neglectable and all of the absorbed light will contribute to an in-phase PA signal. The background signal will in this case be approximately 50 times larger than for the PTFE cell. Due to heat conduction and the insulation properties of the PTFE material the background signal is attenuated further. This has been modeled using a COMSOL simulation applying the heat transfer module, where it was assumed that 15 $\mu$m into the PTFE layer the material is heated by 40 $\mu K$ due to light absorption. From the simulation we find that heating of the surface is "only" approximately 5 $\mu K$. This is due to heat being conducted away via the contact points to the aluminium outer casing of the integrating sphere.

These calculations together with the experimental results demonstrate that reduction of the background signal of typical PA measurements can be addressed by careful choice of the cell wall material in addition to carefully scatter-free optical designs. The heat diffusion and conduction mechanisms suggests that the integrating sphere based PA sensor may be used in the region from 2500 - 3500 nm even though the reflection of the PTFE material is less than $90\%$. This spectral region is very important for trace gas sensing, since the fundamental vibrations of many molecules lie in this spectral region \cite{Sigrist2008,Demtroder2003}.

\section{Conclusion}

The scope of the paper was to demonstrate the PA signal enhancement and the versatility of an integrating sphere based PA sensor as a simple multi-pass absorption cell in combination with an attached organ pipe tube for signal enhancement. A PA enhancement factor of approximately 58(30) for CO$_2$(NO$_2$) in the case of resonant over non-resonant excitation has been demonstrated as the result of attaching a 90 mm long organ pipe tube. Further enhancement has been demonstrated due to the integrating sphere itself resulting in enhancement factors of 20(50) due increased path length. A total enhancement factor of approximately 1200(1500) relative to a simple single pass non-resonant cell with the same incident optical power and microphones has been achieved. The minimum single-shot detectable CO$_2$ concentration in the 2 $\mu$m region is approximately 30 ppm and a minimum detectable NO$_2$ concentration of 1.9 ppm both with a SNR=1 (or SBR=1). It is anticipated that the sensitivity can be further enhanced by applying wavelength modulation techniques, more sensitive microphones, better acoustic insulation, higher optical power and differential measurements, however, this is outside the scope of this paper and will be the focus of the next generation sensor.

It has also been shown that the PA system can be decoupled from various background noise sources, such as the in-phase background absorption signal due to thermal conduction effects and heat diffusion. We believe this is an important result and it demonstrates that the background signal issue of typically PA measurements can be approached from a direction other than through careful scatter-free designs. In the region from 2500 - 3500 nm the reflection of PTFE is less than $90\%$. Since we found that the background absorption signal is greatly attenuated due to thermal conduction of heat away from the absorption cell we believe that the integrating sphere based PA sensor might be used in the region up to 3700 nm. Even though the reflectivity is relatively low. We foresee that this system can be made highly sensitive and versatile at the same time and be very cheap to produce and therefore attract attention as product in the rapidly growing sensor field for climate, environmental and industrial monitoring.\\

\textbf{Acknowledgments}\\
We acknowledge the financial support from the Danish Council for Technology and Production Sciences (Sapere Aude project no. 10-0935849).


\begin{thebibliography}{99}

\bibitem{Sigrist1994} M.W. Sigrist, \emph{Air Monitoring by Spectroscopic Techniques} (John Wiley \& Sons Inc.,1994).

\bibitem{Sigrist2008} M. W. Sigrist, R. Bartlome, D. Marinov, J. M. Rey, D. E. Vogler, and H. Wächter, "Trace gas monitoring with infrared laser-based detection schemes," Appl. Phys. B \textbf{90}, 289--300 (2008).

\bibitem{Patel2008} C. K. N. Patel, "Laser photoacoustic spectroscopy helps fight terrorism: High sensitivity detection of chemical warfare agent and explosives," Eur. Phys. J. Special Topics, \textbf{153}, 1, 1--18 (2008).

\bibitem{Harren2000} F. M. J. Harren, G. Cotti, J. Oomens, and S. te Lintel Hekkert, \emph{Photoacoustic spectroscopy in trace gas monitoring, in Ensyclopedia of Analytical Chemistry} ed. by R. A. Meyers, (John Wiley \& Sons Inc., 2000).

\bibitem{Nägele2000} M. N\"{a}gele and M. W. Sigrist, "Mobile laser spectrometer with novel resonant multipass photoacoustic cell for trace-gas detection," Appl. Phys. B \textbf{70}, 895--901 (2000).

\bibitem{Miklos2001} A. Miklos, P. Hess, and Z. Bozoki, "Application of acoustic resonators in photoacoustic trace gas analysis and metrology," Rev. Sci. Instrum.,\textbf{ 72}, 1937--1955 (2001).

\bibitem{Xu2006} M. Xu, and  L. V. Wang, "Photoacoustic imaging in biomedicine," Review of Scientific Instruments, \textbf{77}, 041101 (2006).

\bibitem{Michaelian2003} K.H. Michaelian, \emph{Photoacoustic Infrared Spectroscopy, Chemical Analysis Series}, ed. by J.D. Winefordner, (John Wiley \& Sons Inc., 2003).

\bibitem{Koskinen2008} V. Koskinen, J. Fonsen, K. Roth and J. Kauppinen, "Progress in cantilever enhanced photoacoustic spectroscopy," Vibr. Spectrosc. \textbf{48}, 1, 16--21, (2008).

\bibitem{Besson2006} J.-P. Besson, S. Schilt, S., L. and Th´evenaz, "Sub-ppm multi-gas photoacoustic sensor," Spectrochim. Acta A, \textbf{63}, 899–-904 (2006).

\bibitem{Rosencwaig1980Book} A. Rosencwaig, \emph{Photoacoustics and Photoacoustic Spectroscopy} (John Wiley \& Sons Inc., 1980).

\bibitem{Webber2003} M. Webber, M. Pushkarsky, and C. Patel, "Fiber-amplifier-enhanced photoacoustic spectroscopy with near-infrared tunable diode lasers," Appl. Opt. \textbf{42}, 2119--2126 (2003).

\bibitem{Rey2005} J. Rey, D. Marinov, D. Vogler, and M. Sigrist, "Investigation and optimisation of a multipass resonant photoacoustic cell at high absorption levels," Appl. Phys. B \textbf{80}, 261–-266 (2005).

\bibitem{Miklos2006} A. Miklos, S.C. Pei, and A.H. Kung, "Multipass acoustically open photoacoustic detector for trace gas measurements," Appl. Opt. \textbf{45}, 2529--2534 (2006).

\bibitem{Saarela2010} J. Saarela, Johan Sand, T. Sorvajarvi, A. Manninen, and J. Toivonen, "Transversely Excited Multipass Photoacoustic Cell Using Electromechanical Film as Microphone," Sensors \textbf{10}, 5294--5307 (2010).

\bibitem{Manninen2012}  A. Manninen, B. Tuzson, H. Looser, Y. Bonetti, and L. Emmenegger, "Versatile multipass cell for laser spectroscopic trace gas analysis," Applied Physics B \textbf{109}, 3, 461--466 (2012).

\bibitem{Elterman1970} P. Elterman, "Integrating Cavity Spectroscopy Applied Optics," Vol. \textbf{9}, 2140--2142 (1970).

\bibitem{Hodgkinson2009} Jane Hodgkinson, Dackson Masiyano, and Ralph P. Tatam, "Using integrating spheres as absorption cells: path-length distribution and application of Beer's law", Applied Optics, Vol. \textbf{48}, 30, 5748--5758 (2009).

\bibitem{Tranchart1996} S. Tranchart, I. H. Bachir, and J.-L. Destombes, "Sensitive trace gas detection with near-infrared laser diodes and an integrating sphere," Appl. Opt. \textbf{35}, 7070–-7074 (1996).

\bibitem{Hawe2005} E. Hawe, E. Lewis, and C. Fitzpatrick, "Hazardous gas detection with an integrating sphere in the near-infrared," J. Phys. Conf. Ser. \textbf{15}, 250–-255 (2005).

\bibitem{Lewicki2007} R. Lewicki, G. Wysocki, A. A. Kosterev, and F. K. Tittel, "Carbon Dioxide and ammonia detection using 2µm diode laser based quartz-enhanced photoacoustic spectroscopy," Appl. Phys. B \textbf{87}, 157--162 (2007).

\bibitem{Hawe2006} E. Hawe, G. Dooly, C. Fitzpatrick, E. Lewis, and P. Chambers, "UV based pollutant quantification in automotive exhausts," Proc. SPIE \textbf{6198}, 619807 (2006).

\bibitem{Bernhardt2010} R. Bernhardt, G. D. Santiago, V. B. Slezak, A. Peuriot, and M. G. Gonzlez, "Differential, LED-excited, resonant NO$_2$ photoacoustic system," Sens. Actuators B \textbf{150}, 513–-516 (2010).

\bibitem{Yi2011} H. Yi, K. Liu, W. Chen, T. Tan, L. Wang, and X. Gao, "Application of a broadband blue laser diode to trace NO$_2$ detection using off-beam quartz-enhanced photoacoustic spectroscopy," Opt. Lett. \textbf{36}, 481–-483 (2011).

\bibitem{PAS1881} A. G. Bell, Phil. Mag.,"The production of sound by radiant energy," \textbf{11}, 510 (1881).

\bibitem{Tam1986} A. C. Tam, "Applications of photoacoustic sensing techniques," Reviews of Modern Physics, \textbf{58}, 381, (1986).

\bibitem{Demtroder2003}  W. Demtroder, \emph{Laser Spectroscopy: Basic Concepts and Instrumentation} third Edition, (Springer-Verlag, 2003).

\bibitem{Harren1997} F. Harren, and Jorg Reuss, \emph{Photoacoustic Spectroscopy}, ed. G.L. Trigg,  (Wiley-VCH Verlag GmbH, 1979).

\bibitem{Schilt2006} S. Schilt and L. Thevenaz, "Wavelength modulation photoacoustic spectroscopy: Theoretical description and experimental results," Infrared Phys. Technol. \textbf{48}, 154--162 (2006).

\bibitem{Pitts1964} J. N. Pitts Jr., J. H. Sharp, and S. I. Chan,"Effects of Wavelength and Temperature on Primary Processes in the Photolysis of Nitrogen Dioxide and a Spectroscopic—Photochemical Determination of the Dissociation Energy," J. Chem. Phys. \textbf{40}, 3655, (1964)

\bibitem{Barreiro2010} N. Barreiro, A. Vallespi, A. Peuriot, V. Slezak, and G. Santiago, "Quenching effects on pulsed photoacoustic signals in NO$_2$-air samples," Appl. Phys. B: Lasers Opt. \textbf{99}, 591–-597 (2010)

\bibitem{Saarela2011}J. Saarela, T. Sorvaj\"{a}rvi, T. Laurila, and J. Toivonen, "Phase-sensitive method for background-compensated photoacoustic detection of NO$_2$ using high-power LEDs", Optics Express, Vol. \textbf{19}, 725--732 (2011)

\bibitem{Sidorov2012} I. S. Sidorov, S. V. Miridonov, E. Nippolainen, and A. A. Kamshilin, "Estimation of light penetration depth in turbid media using laser speckles," Optics Express, \textbf{20}, 13, 13692--13701 (2012)

\bibitem{Bird1976} R.B. Bird, W.E. Stewart and E.N. Lightfoot, \emph{Transport Phenomena} (John Wiley \& sons, 1976)

\end{thebibliography}
\end{document}